\documentclass[pra,aps,twocolumn,superscriptaddress,showpacs]{revtex4-1}
\usepackage[T1]{fontenc}
\usepackage[latin9]{inputenc}
\usepackage{geometry}
\usepackage{amssymb}
\usepackage{graphicx}
\usepackage{esint}
\usepackage{array}
\usepackage{multirow}
\usepackage{color} %for change tracking

\newcommand{\ket}[1]{\ensuremath{\left|#1\right\rangle}}

\begin{document}

\title{Generation of tunable entanglement and violation of a Bell-like inequality between different degrees of freedom of a single photon}

\author{Adam Vall\'es} \email{adam.valles@icfo.es}
\affiliation{ICFO---Institut de Ci\`encies Fot\`oniques, Mediterranean
Technology Park, 08860, Castelldefels, Barcelona, Spain}

\author{Vincenzo D'Ambrosio}
\affiliation{Dipartimento di Fisica, Sapienza Universita di
Roma, Roma 00185, Italy }

\author{Martin Hendrych}
\affiliation{ICFO---Institut de Ci\`encies Fot\`oniques, Mediterranean
Technology Park, 08860, Castelldefels, Barcelona, Spain}

\author{Michal Mi\v cuda}
\affiliation{Department of Optics, Palack\'y University, 17 listopadu 12,
77146 Olomouc, Czech Republic}

\author{Lorenzo Marrucci}
\affiliation{Dipartimento di Fisica, Universita di Napoli Federico
II, Complesso Universitario di Monte Sant'Angelo, Napoli, Italy}

\author{Fabio Sciarrino}
\affiliation{Dipartimento di Fisica, Sapienza Universita di Roma,
Roma 00185, Italy }

\author{Juan P. Torres}
\affiliation{ICFO---Institut de Ci\`encies Fot\`oniques, Mediterranean
Technology Park, 08860, Castelldefels, Barcelona, Spain}
\affiliation{Department of Signal Theory and Communications,
Universitat Polit\`ecnica de Catalunya, Jordi Girona 1-3, Campus Nord D3, 08034
Barcelona, Spain}

\begin{abstract}
We demonstrate a scheme to generate noncoherent and coherent
correlations, i.e., a tunable degree of entanglement,  between
degrees of freedom of a single photon. Its nature is analogous to
the tuning of the purity (first-order coherence) of a single
photon forming part of a two-photon state by tailoring the
correlations between the paired photons. Therefore, well-known
tools such as the Clauser-Horne-Shimony-Holt (CHSH) Bell-like
inequality can also be used to characterize entanglement between
degrees of freedom. More specifically, CHSH inequality tests are
performed, making use of the polarization and the spatial shape of
a single photon. The four modes required are  two polarization
modes and two spatial modes with different orbital angular
momentum.
\end{abstract}


\maketitle

\section{Introduction}
Entanglement, a concept introduced in quantum theory nearly eighty
years ago by Schr\"{o}dinger \cite{schrodinger1935}, is one of
the main traits of quantum theory; for some it is even its {\em
weirdest feature} \cite{weinberg2013}. Since the publication of
the seminal gedanken experiment by Einstein, Podoslky, and
Rosen (EPR) in their famous 1935 paper \cite{EPR1935} and the
appearance of the first comments about it the very same year
\cite{bohr1935}, innumerable theoretical discussions and
experiments related to this subject have appeared.

Arguably the most relevant contribution to this discussion has
been the introduction, now fifty years ago, of the now
well-known Bell inequalities \cite{bell1964}. One of these
Bell-like inequalities, the Clauser-Horne-Shimony-Holt (CHSH)
inequality \cite{CHSH1969}, which will be used in this work, is the
most commonly used one in experiments \cite{aspect1982}.
Originally, Bell's inequalities were considered for composite
systems made up of two separate subsystems, i.e., two subsystems
propagating along different directions that had interacted in the
past.

For instance, the two subsystems can be each one of the two
photons generated by means of the nonlinear process of spontaneous
parametric down-conversion (SPDC), when an intense pump beam
interacts with the atoms of a noncentrosymmetric nonlinear crystal
\cite{torres2011}. Entanglement can reside in any of the degrees
of freedom that characterize each of the photons, with being
polarization the most common. In this case, one of the quantum
states that allows a maximum violation of the CHSH inequality can
be written as $|\Phi\rangle=1/\sqrt{2}\, \left[
a_{k_1,H}^{\dagger} a_{k_2,V}^{\dagger}+a_{k_1,V}^{\dagger}
a_{k_2,H}^{\dagger}\right] |\text{vac} \rangle$, where $a_{k_i,H}^{\dagger}$ designates the creation operator of a photon propagating
along direction $k_i$ ($i=1,2$) with polarization $H$, similarly for $a_{k_i,V}^{\dagger}$, and $|\text{vac} \rangle$ is the vacuum state.

\begin{figure*}[t]
\centering
\includegraphics[width=15.9cm]{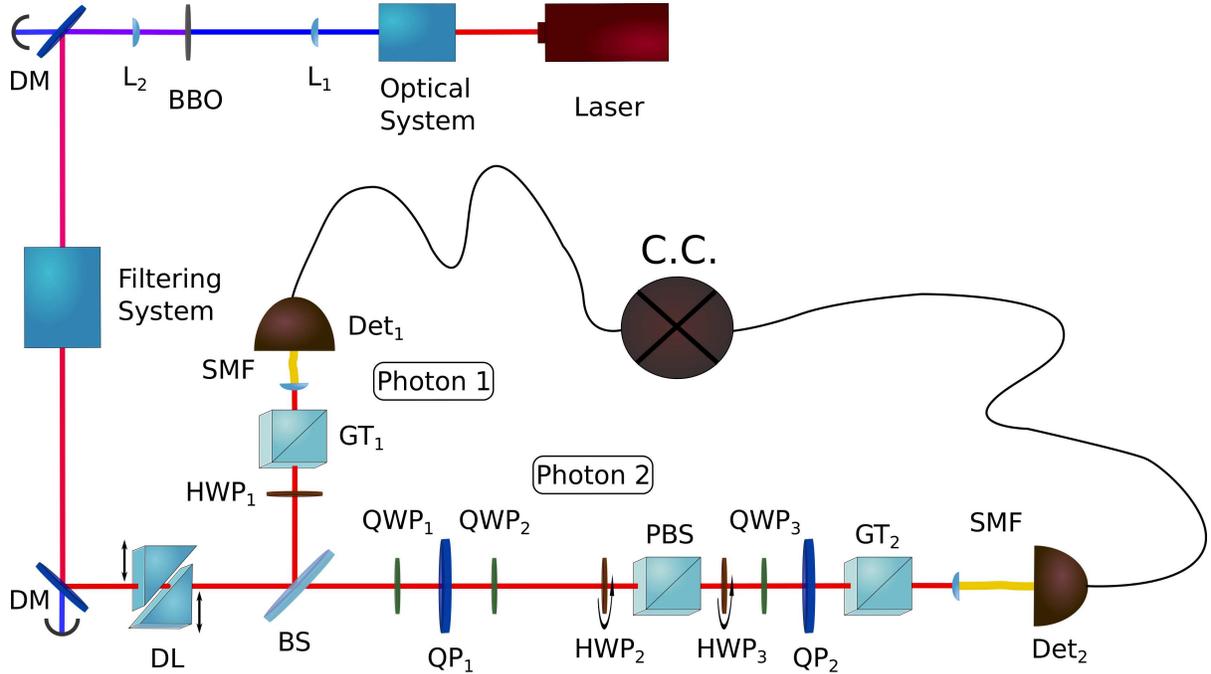}
\caption{(Color online) Experimental setup scheme. Laser: Mira
900 (Coherent). Optical system: second harmonic generation
(Inspire Blue, Radiantis), spatial filter, linear attenuator, three
dichroic mirrors (DM), and short pass filter. L$_1$ and L$_2$:
Fourier lenses. BBO: nonlinear crystal. Filtering system:
long-pass and band-pass filters. DL: delay line. BS: beam splitter (50:50). HWP$_1$, HWP$_2$,
and HWP$_3$: half-wave plates. PBS: polarization beam splitter.
GT$_1$ and GT$_2$: Glan-Thompson polarizers. QWP$_1$, QWP$_2$, and
QWP$_3$: quarter-wave plates. QP$_1$ and QP$_2$: q-plates. Det$_1$
and Det$_2$: single-photon counting modules. C.C.:
coincidence-counting electronics.} \label{fig1}
\end{figure*}

However, correlations of a nature similar to the ones existing
between physically separated photons can also exist considering
different degrees of freedom of a single system. Therefore, Bell's
inequalities can be used as well to characterize these
correlations existing between different parts of a single system.
The key point to consider regarding Bell's inequalities in this scenario is
the capability to perform independent measurements in any of the
degrees of freedom involved. In Ref. \cite{gadway2009}, a single photon was generated in the quantum state $|\Phi\rangle=1/\sqrt{2}\, [
a_{k_1,H}^{\dagger} + a_{k_2,V}^{\dagger} ] |\text{vac} \rangle$,
which violates a Bell-like inequality involving two degrees of
freedom (polarization and path).

Bell-like inequalities can be also used to characterize beams
containing many photons, i.e., intense beams, coherent or not. In
Refs. \cite{borges2010,kagalwala2013}, the authors make use of coherent
beams whose electric field reads ${\bf E}({\bf r})=1/\sqrt{2}\, [
\Psi_H({\bf r}) \hat{\bf e}_H  + \Psi_V({\bf r}) \hat{\bf e}_V]$
and use a CHSH inequality to characterize their coherence
properties in one of the two degrees of freedom involved, i.e.,
polarization or the spatial shape. Entanglement, as the
inseparability of degrees of freedom, has also been considered
\cite{simon2010,eberly2011} as a fundamental tool to address and
shed new light on certain characteristics of classical fields,
by applying analysis and techniques usually restricted to
entanglement in a quantum scenario.

Here we intend to move further into this analogy and show
experimentally that one can generate tunable entanglement between
two degrees of freedom of a single photon, going from the
generation of coherent correlations to incoherent ones. For the
single-photon case, the control of the degree of entanglement
between degrees of freedom is fully equivalent to tuning the
first-order coherence \cite{glauber1966} of one of the degrees of
freedom involved, in full analogy with the relationship existing
between the degree of entanglement between separate photons and
the first-order coherence of one of the photons that forms the
pair.

Different types of quantum states provide different results in the
measurement of the CHSH inequality. This notwithstanding, for any
quantum state with any degree of first-order coherence or purity,
we demonstrate that the results of a Bell's measurement obtained
using different degrees of freedom of a single photon, are the
same as when using the properties of separate photons.

In our experiments we make use of single photons where the two
degrees of freedom involved are the polarization (horizontal and
vertical linear polarizations) and spatial modes (two spatial
modes with orbital angular momentum index  $m=\pm 1$). The orbital
angular momentum (OAM) states allow for a relatively simple
experimental generation, filtering, detection, and control
\cite{nature}. These states are characterized by the index $m$,
which can take any integer number, and determines the azimuthal
phase dependence of the mode, which is of the form $\sim \exp
\left( i m \varphi\right)$. Each mode carries an OAM of $m\hbar$
per photon. The feasibility to generate entangled states in the
laboratory using polarization and spatial modes with OAM is
greatly facilitated by the use of the so-called q-plates
\cite{marrucci2006}: Liquid crystal devices which couple together
polarization and orbital angular momentum and allow the generation
of states that have been recently exploited in fundamental quantum
mechanics \cite{nagali2012,dambrosio2013}, quantum communications
\cite{dambrosio2012}, and metrology \cite{dambrosio2013gear}. In
Ref. \cite{nagali2009}, Nagali et al. generated a single-photon quantum
state with the OAM and polarization degrees of freedom with high
purity. Karimi et al. \cite{karimi2010} used this same state to
demonstrate the violation of the CHSH inequality.

\section{Experimental setup}
The experimental setup used in our experiments is shown in
Fig.~\ref{fig1}. Paired photons are generated in a $2$-mm-long
$\beta$-barium borate (BBO) nonlinear crystal by means of spontaneous
parametric down-conversion (SPDC). We choose a type-II source,
where the photons generated have orthogonal (horizontal and
vertical) polarizations in order to generate a
polarization-entangled photon pair by postselection with a beam
splitter and a coincidence detection.

\begin{figure}[t]
\begin{center}
\includegraphics[width=9.2cm]{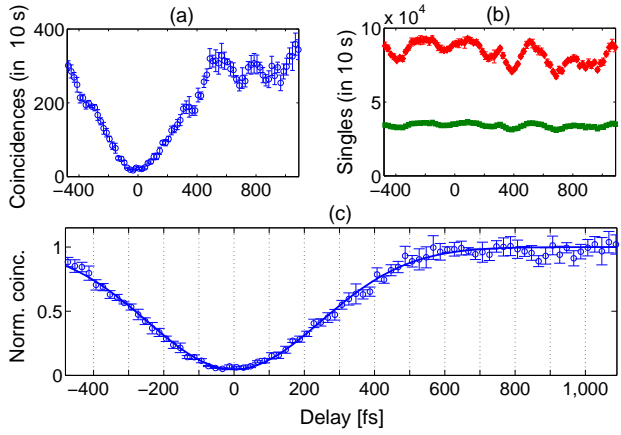}
\end{center}
\caption{(Color online) Coincidence and singles detections as a function of the
temporal delay $\tau$ in a Houng-Ou-Mandel (HOM) interferometer. The raw data of the
coincidences measured in 10 s are plotted in panel (a), and the
singles detected for each detector are shown in panel (b). Closed
diamonds (upper curve) correspond to singles detected with
detector $1$, and closed squares (lower curve) correspond
to measurements in detector $2$. The compensated and normalized
number of coincidences is plotted in panel (c), using the coincidence
data of panel (a) and the singles detected with detector $1$ shown in
panel (b).} \label{fig3}
\end{figure}

The pumping laser is a Mira 900 (Coherent) working in the
picosecond regime and tuned to a central wavelength of $810$ nm.
In order to obtain the down-converted photons at $810$ nm, light
from Mira is frequency doubled in a second-harmonic setup
(Inspire Blue, Radiantis). The output light at $405$ nm traverses
an optical system with five dichroic mirrors and a short-pass
filter to filter out the remaining $810$-nm light. A spatial
filter tailors the spatial shape of the pump beam to obtain the
sought-after Gaussian beam profile. We use a $750$-mm focal
distance lens to obtain a pump beam with $400$-$\mu$m beam waist
that is focused in the middle of the nonlinear crystal. A smaller
beam waist would increase efficiency of the SPDC
process; however, the spatial walkoff in the BBO crystal impedes
tighter focusing, because it would also introduce harmful spatial
distinguishability between the generated photons. The
down-converted photons are collimated with a $400$-mm focal
distance lens.

Another filtering system, formed by two dichroic mirrors, a
long-pass filter, and a band-pass filter, removes the residual pump
light at $405$ nm. Different group velocities result in slightly
different spectra of the orthogonal polarizations, thus mixing the
polarization and frequency properties of the photons. The use of a
filter with $3$-nm full-width-half-maximum bandwidth centered at
$810$ nm helps reducing the spectral distinguishability between
the photons.

\begin{figure}[t]
\centering \includegraphics[width=9.2cm]{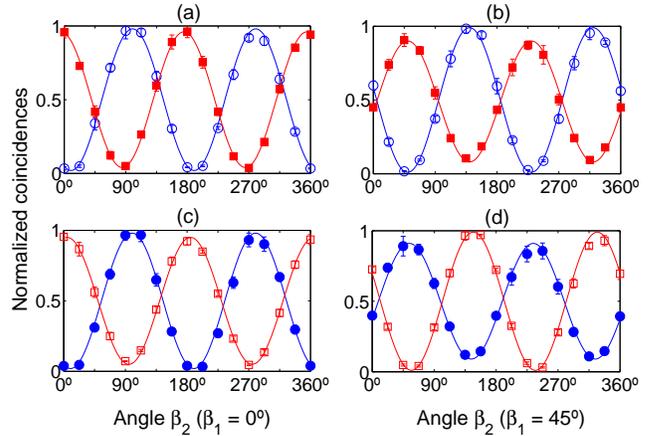}
\caption{(Color online) Normalized value of the coincidences as a function of the
projection angle $\beta_2$. Panels (a) and (c): the angle of HWP$_2$ is
set to $\beta_1=0^\circ$; panels (b) and (d): the angle is set to
$\beta_1=45^\circ$. Curves corresponding to experimental values
are shown with error bars. Solid lines are theoretical
predictions. Open circles, $\ket{\Psi^-}$;  closed circles,
$\ket{\Psi^+}$; open squares, $\ket{\Phi^-}$; and closed squares,
$\ket{\Phi^+}$.} \label{fig2}
\end{figure}

After the beam splitter, the quantum state of the two photons,
considering only the cases when the paired photons are detected in
coincidence (postselection), can be generally written as
\begin{eqnarray}
\label{state_after_BS}
 & & \rho=\epsilon  |\Psi\rangle \langle \Psi|+ \frac{1-\epsilon}{2} \nonumber \\
 & & \times \left\{ |H \rangle_{1} |V \rangle_{2} \langle H |_1 \langle V |_2 + |V \rangle_{1} |H \rangle_{2}  \langle V |_1 \langle H |_2\right\},
\end{eqnarray}
where indexes $1$ and $2$ refers to paths $1$ and $2$
after the beam splitter, $|\Psi\rangle=1/\sqrt{2}\, [|H
\rangle_{1} |V \rangle_{2}+|V \rangle_{1} |H \rangle_{2}]$, and
$\epsilon$ depends on the delay ($\tau$) between the two
orthogonal photons generated. The form of the state given by Eq.~(\ref{state_after_BS}) is due to the correlation exiting between
the polarization of the photon generated and its group velocity,
since the nonlinear crystal used (BBO) is a birefringent crystal.
In particular, the group velocity of photons at $810$ nm with
horizontal polarization (ordinary wave) is $v_g^o=1.7816 \times
10^{8}$ m/s, while the group velocity of photons with vertical
polarization (extraordinary wave) is $v_g^e=1.8439 \times 10^{8}$
m/s, which produces a group velocity mismatch (GVM) of
$D_{BBO}=1/v_g^o-1/v_g^e=189.6$ fs/mm. This distinguishability of
photons by its group velocity cause the mixed character of the
quantum state in polarization given by Eq.~(\ref{state_after_BS}).

A delay line, formed by quartz prisms, can be used to tune its
value. If photons could be distinguished by their time of arrival
at the detectors, then $\epsilon=0$ and the purity of the quantum
state that describes the two photons generated is minimal
($P=1/2$). The purity of the quantum state can be increased by
adding or removing the length of quartz that the photons traverse
along its optical path \cite{martin2012}, which is necessary to remove all
distinguishing information coming from the temporal-frequency
degree of freedom. The group velocity of ordinary waves in quartz
is $v_g^o=1.9305 \times 10^{8}$ m/s, while the group velocity of
extraordinary waves is $v_g^e=1.9187 \times 10^{8}$, which produces
a GVM of $D_{quartz}=-31.8$ fs/mm. For a
specific arrangement of the quartz prisms, that we define as
$\tau=0$, we can have $\epsilon=1$. For the $L=2$ mm long BBO
crystal of our experiment, with group velocity mismatch of
$D_{BBO}=189.6$ fs/mm, this requires \cite{kwiat1995} compensating
with the tunable delay line $D_{\small BBO}L/2=189.6/2$ fs/mm
$\times$ 2 mm =$189.6$ fs.

To entangle the polarization and the orbital angular momentum
(OAM) degrees of freedom in a single photon, the photon reflected
from the the beam splitter (photon $1$) is projected into the
linear diagonal polarization state: $1/\sqrt{2}\, \left[|H\rangle
\pm|V\rangle\right]$, with a half-wave plate (HWP$_1$) and a
Glan-Thompson polarizer (GT$_1$), coupled into a single mode
fiber, to remove the remaining spatial distinguishability
introduced by the presence of spatial walkoff in the BBO crystal,
and detect it in coincidences (coincidence time window of
12.5 ns). The transmitted photon (photon $2$) traverses a
quarter-wave plate (QWP$_1$) to rotate its polarization from
horizontal and vertical to circular right ($\bf R$) and circular left
($\bf L$), a q-plate (QP$_1$) correlates polarization with OAM,
and another quarter-wave plate (QWP$_2$) transforms the
polarization back from circular right and circular left to
horizontal and vertical. In summary,
\begin{eqnarray}
& & |\bf H \rangle \Longrightarrow |\bf R \rangle \Longrightarrow
|\bf L,m=-1 \rangle \Longrightarrow |\bf H, m=-1 \rangle,
\nonumber
\\
& & |\bf V \rangle \Longrightarrow |\bf L \rangle \Longrightarrow
|\bf R,m=+1 \rangle \Longrightarrow |\bf V, m=+1 \rangle.
\end{eqnarray}
After the second quarter-wave plate, the quantum state of photon
$2$, after projection and detection of photon $1$, is written as
\begin{eqnarray}
\label{state_experiment}
 & & \rho=\epsilon  |\Psi^{\pm}\rangle \langle \Psi^{\pm}|+ \frac{1-\epsilon}{2} \left[ |H,m=-1 \rangle \langle H, m=-1| \right. \nonumber \\
 & & \left. +|V,m=+1 \rangle \langle V, m=+1|  \right],
\end{eqnarray}
where
\begin{equation}
\ket{\Psi^{\pm}} \frac{1}{\sqrt{2}} \left(\ket{H,m=-1} \pm
\ket{V,m=+1} \right). \label{Bellstate1}
\end{equation}
The purity of the state is $P=(1+\epsilon^2)/2$. If one would
apply the concept of concurrence \cite{wootters1998} to this
single-photon state, considering as the two subsystems the
polarization and OAM degrees of freedom of the photon, one would
obtain $C=\epsilon$.

\begin{figure}[t!]
\centering \includegraphics[width=9.4cm]{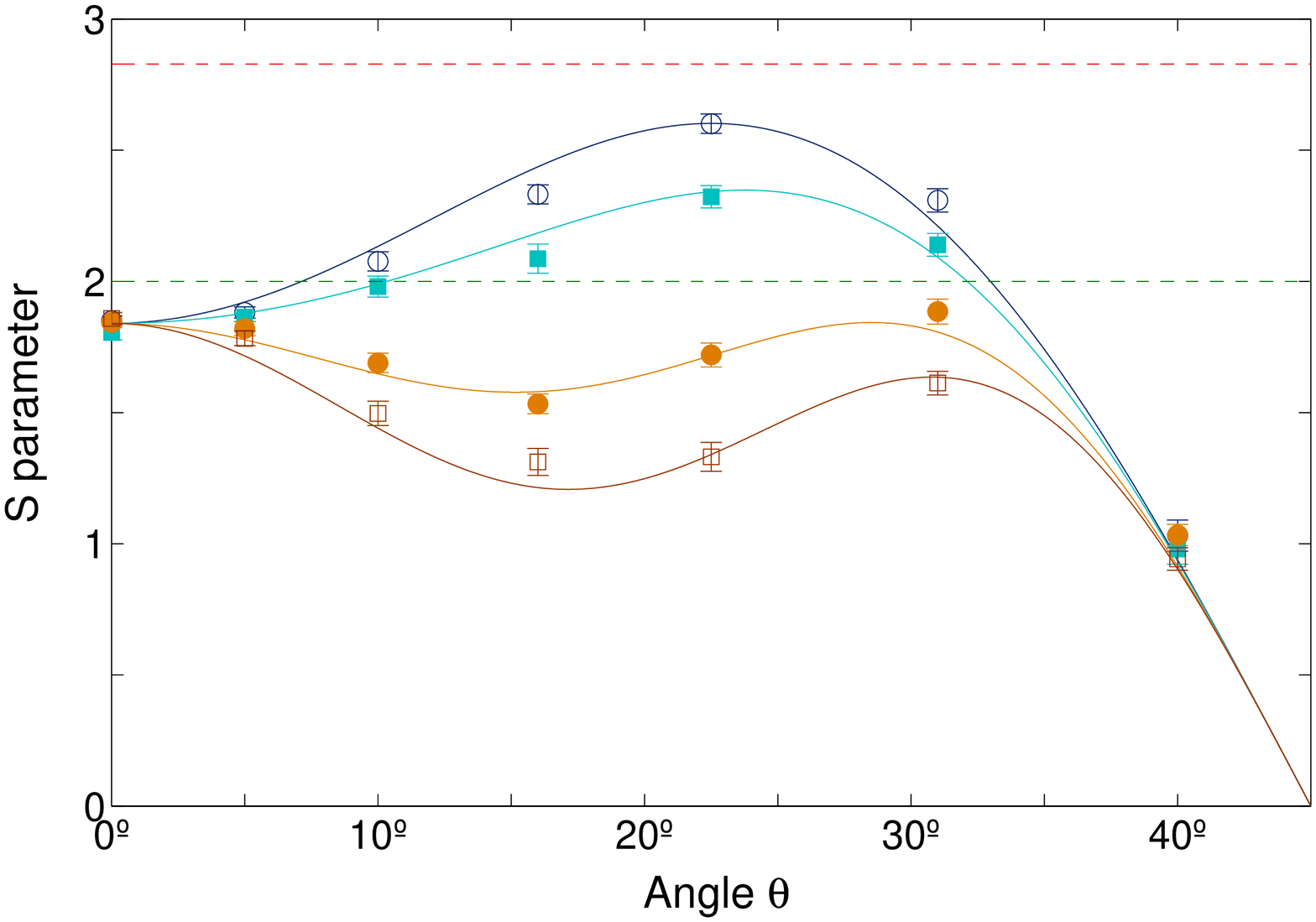}
\caption{(Color online) Value of the parameter $S$ in a CHSH
inequality as a function of the angle $\theta=b_1-a_1$. The colored
symbols with error bars represent the experimental data with
their standard deviations. The solid colored curves are the
theoretical predictions assuming a visibility factor of $V=0.92$.
Open circles, $\epsilon=1$; closed squares, $\epsilon=0.8$;
closed circles, $\epsilon=0.32$; and open squares,
$\epsilon=0.03$. The values of $\epsilon$ correspond to delays
of 0 fs ($\epsilon=1$), 200 fs ($\epsilon=0.8$), 400 fs ($\epsilon=0.32$), and 600 fs ($\epsilon=0.03$), as depicted in the HOM dip of
Fig.~\ref{fig3}. The dashed red line (upper) corresponds to the
Tsirelson bound, and the dashed green line (lower) is the CHSH
inequality limit.} \label{fig4}
\end{figure}

The measurement stage consist of  projecting the quantum state
generated into specific polarization and OAM states in two steps.
First, the state of polarization is projected into the desired
state with a half-wave plate (HWP$_2$) and a polarizing beam
splitter (PBS). The OAM can be projected into any state using
several polarization optic elements, before and after a second
q-plate (QP$_2$) \cite{nagali2009}. More specifically,  the OAM
state information is transferred into a polarization state with a
half-wave plate (HWP$_3$) and a quarter-wave plate (QWP$_3$)
located before the q-plate, to transform horizontal-vertical
polarizations to right-left polarizations base, and another
Glan-Thompson polarizer (GT$_2$) located after . Finally, the
photon is spatially filtered by coupling it to a single-mode fiber
and detecting it in coincidence with the other photon.

\section{Experimental results}
In order to be able to relate the value of $\epsilon$ in Eq.~(\ref{state_experiment}) to the delay introduced by the delay
line, and determine the value of the delay which makes the quantum
state pure ($\epsilon=1$), we construct a Hong-Ou-Mandel
interferometer (HOM). If we choose the temporal delay introduced
by the delay line so that coincidences are close to zero, the
state given by Eq.~(\ref{state_experiment}) is pure ($\epsilon=1$)
and corresponds to a Bell state. We choose to generate the quantum
state $\ket{\Psi^{-}}$ to obtain the HOM dip. Figure~\ref{fig3}(a)
shows the coincidence photons measured in detectors 1 and 2, and
Fig.~\ref{fig3}(b) shows the single photons detected in each
detector.
Figure~\ref{fig3}(c) shows coincidence detections renormalized
using the single measurements from detector $1$. The oscillations
in detector 1 are due to imperfections in the translation stage of
the delay line (DL), causing deviations in the photon
trajectories. Thus the single detections  of detector 1 are
clearly affected by these corresponding variations in the coupling
efficiency. We should notice that all the results presented in this
paper are shown with no substraction of the accidental
coincidences ($\sim 4$ pairs in $10$ s).

\begin{figure}[t!]
\centering
\includegraphics[width=9.2cm]{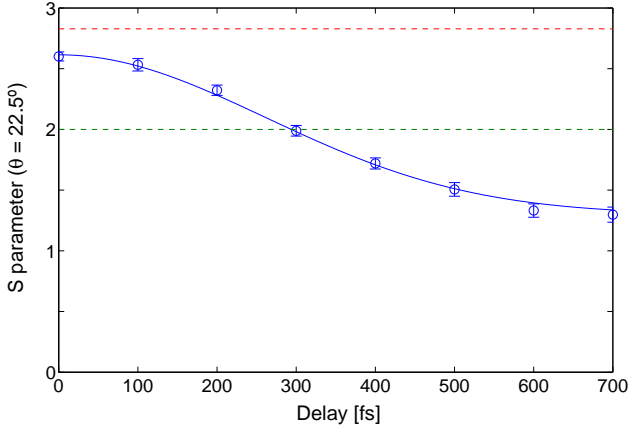}
\caption{(Color online) Value of the CHSH inequality
for $\theta=22.5^\circ$ as a function of the temporal delay, as
depicted in Fig.~\ref{fig3}. The solid (blue) curve is the theoretical prediction assuming a visibility factor of $V=0.92$. The dashed red line (upper)
corresponds to the Tsirelson bound ($S_{max}=2\sqrt{2}$), and the
dashed green line (lower) is the CHSH inequality limit ($S=2$).}
\label{fig5}
\end{figure}

When we change the projection of  photon $1$ from the state
$1/\sqrt{2}\, \left[ \ket{H}+\ket{V}\right]$ to $1/\sqrt{2}\,
\left[ \ket{H}-\ket{V}\right]$ with HWP$_1$, we change the sign of
the corresponding Bell state, from $\ket{\Psi^{-}}$ to
$\ket{\Psi^{+}}$. By modifying the transformation of photon $2$
from $L/R \Longrightarrow H/V$ to $L/R \Longrightarrow V/H$ with
QWP$_2$, we can go from the generation of $\ket{\Psi^\pm}$ to
$\ket{\Phi^\pm}$, where $\Phi^{\pm}$ can be written as
\begin{equation}
\ket{\Phi^{\pm}} = \frac{1}{\sqrt{2}}\left(\ket{H,m=+1} \pm \ket{V,m=-1} \right). \nonumber \\
    \label{Bellstates}
\end{equation}
With this procedure we are able to create the four Bell states.

Figure~\ref{fig2} shows the coincidences measured for each of the
four Bell states. Photon $2$ is projected first into the
polarization state $\sim \cos \beta_1 \ket{H} + \sin\beta_1
\ket{V}$, with $\beta_1=0^{\circ}$,$45^{\circ}$, and after that a
second projection is performed into a set of OAM states of the
form $\cos \beta_2 \ket{+1} + \sin \beta_2 \ket{-1}$, with
$\beta_2$ spanning from $0$ to $2\pi$. Ideally, for the state
$\ket{\Psi^-}$, coincidence counts as a function of $\beta_2$
follow the form of $\sin^2(\beta_1-\beta_2)$, which yields a
visibility \cite{kwiat1995} V=(max-min)/(max+min) of $100\%$.
Therefore, as the visibility measured increased, so did the
quality of the entangled state generated. The small phase shifts
observed in the curves are due to some misalignment still present
between the position of the centers of the vortex of the two OAM
modes, $m=+1$ and $m=-1$, when going through the second q-plate
(QP$_2$).

Measurements of the CHSH inequality \cite{CHSH1969}  requires
choosing two polarization states and two OAM states where the
state of photon $2$, given by Eq.~(\ref{state_experiment}), is
projected. When considering any possible state projection,
following Ref. \cite{horodecki1995}, one finds that the maximum
violation of the CHSH inequality for this state is
\begin{equation}
S_{max}=2\sqrt{1+\epsilon^2}.
\end{equation}
For $\epsilon=1$ we reach the Tsirelson bound. We will restrict the discussion here to only projections into states of the form
\begin{eqnarray}
    & & \ket{{\bf a}_i} = \frac{1}{\sqrt{2}}\left(\cos a_i \ket{H}+\sin a_i \ket{V} \right),\nonumber \\
    & & \ket{{\bf b}_i} = \frac{1}{\sqrt{2}}\left(\cos b_i \ket{m=+1} + \sin b_i
    \ket{m=-1} \right),
    \label{projections}
\end{eqnarray}
where states ${\bf a}_i$ ($i=1,2$) refers to linear polarization
states and  ${\bf b}_i$ ($i=1,2$) refer to OAM states which are
linear combinations of modes $m=+1$ and $m=-1$. By proper
combinations of all of the polarization optical elements of the
setup (half-wave and quarter-wave plates), one can project the
photon into any combination (${\bf a}_i, {\bf b}_i$) as required.

For the single-photon case, restricting our attention to state
projections of the form given in Eq.~(\ref{projections}), the CHSH
inequality can be written as
\begin{equation}
S=E(a_1,b_1)-E(a_1,b_2)+E(a_2,b_1)+E(a_2+b_2) \le 2,
\end{equation}
where
\begin{widetext}
\begin{equation}
E(a_i,b_i)=\frac{N_{++}(a_i,b_i)+N_{--}(a_i^{\perp},b_i^{\perp})-N_{+-}(a_i,b_i^{\perp})-N_{-+}(a_i^{\perp},b_i)}{N_{++}(a_i,b_i)+N_{--}(a_i^{\perp},b_i^{\perp})+N_{+-}(a_i,b_i^{\perp})+N_{-+}(a_i^{\perp},b_i)}.
\end{equation}
\end{widetext}
$N_{++}(a_i,b_i)$ is the number of photons detected when its
quantum state is projected into a polarization state determined by
the angle $a_i$ and an OAM state determined by the angle $b_i$.
All other cases follow similarly, taking into account that
$a_i^{\perp}=a_i+\pi/2$ and $b_i^{\perp}=b_i+\pi/2$. One can
easily find that for the state given by Eq.~(\ref{state_experiment}),
\begin{equation}
E(a_i,b_i)=\cos 2a_i \cos 2 b_i+\epsilon \sin 2a_i \sin 2 b_i.
\end{equation}

Figure~\ref{fig4} shows the value of $S$ measured when we go from
a pure to a mixed state, i.e., for different values of $\epsilon$
from $0$ to $1$. It shows the value of $S$ as a function of the
angle $\theta$, where $\theta \equiv b_1 - a_1 = b_2 + a_2 = - b_1
- a_2$. For the case of a pure state, one would obtain $ S(\theta)
= 3\cos 2\theta - \cos 6\theta$. The experimental values measured
decrease from the theoretical (ideal) expected values due to the
existence of accidental coincidences or the inevitable
misalignment of optical elements, by a factor $V$, the visibility
measured in Fig.~\ref{fig2}. In our case, the maximum CHSH
inequality value measured is $S(\theta=22.5^\circ) = 2.601 \pm
0.037$ and the visibility is $V=0.92$.

Figure~\ref{fig4} shows that there is a complete analogy between a
Bell-like inequality involving the same degree of freedom of two
separate photons \cite{aspect1982,horodecki1995} and that involving two distinct degrees of
freedom of the same single photon, independent of the purity (or
first-order coherence) of the quantum state. Figure~\ref{fig5}
shows the CHSH violation measured for $\theta=22.5^{\circ}$, which
gives the maximum violation for a pure state. When the delay
increases or decreases from $\tau_0$, the state becomes
increasingly mixed and entanglement disappears. Figures~\ref{fig4}
and \ref{fig5} are very similar to what would have been obtained
for the case of two separate correlated photons, even though here
the measurement corresponds to measuring correlations between
properties in different degrees of freedom of a single photon. The
similarities in form between the quantum states with different
numbers of photons are why we obtain similar results,
as it has been pointed out in several theoretical papers
\cite{simon2010,eberly2011} and experiments \cite{gadway2009,borges2010}.

\section{Conclusions}
In conclusion, we have demonstrated experimentally that there is a
full analogy between the general quantum state (pure or mixed)
that describes two-photon states entangled in its polarization
degree of freedom and the correlations (coherent or noncoherent)
existing between the polarization and spatial degrees of freedom
of a single photon. Along these lines, concepts such as purity and
degree of entanglement or concurrence can be used to
describe coherent and noncoherent correlations between
properties of a single system. This fact naturally allows one to
use Bell's inequalities to characterize both types of systems, as
we have demonstrated here.

\begin{acknowledgments} We thank Fabrizio Bisesto for his collaboration in the first stage of the experiment. This work was supported by the project FIS2010-14831, the program Severo Ochoa, from the
Government of Spain, and the FET-Open Program, within the
Seventh Framework Program of the European Commission under
Grant No. 255914 (PHORBITECH). This work has also been partially
supported by Fundacio Privada Cellex Barcelona. V.D'A. and F.S. acknowledge support via the Starting Grant 3D-QUEST (3D-Quantum Integrated Optical Simulation; Grant Agreement No. 307783). M.M. acknowledges support by the European Social Fund and MSMT
under Project No. EE2.3.20.0060.
\end{acknowledgments}

\end{document}